\documentclass[10pt,aps,pra,twocolumn,showpacs,superscriptaddress]{revtex4-2} 

\usepackage[utf8]{inputenc}
\usepackage{graphicx}
\usepackage{caption}
\usepackage{amsmath}
\usepackage{amssymb}
\usepackage{bm}
\usepackage{txfonts}
\usepackage{multirow}
\usepackage{mathptmx} 
\usepackage{siunitx}
\usepackage{xargs}                    
\usepackage[pdftex,dvipsnames]{xcolor}
\usepackage[colorlinks=true, urlcolor=blue, linkcolor=blue, citecolor=blue]{hyperref}
\usepackage{xcolor}
\usepackage{braket}
\usepackage{subfig}
\usepackage{soul}
\begin{document}

\title{Synthesizing a $\hat{\sigma}_z$ spin-dependent force for optical, metastable, and ground state trapped-ion qubits}
\author{O. B\u{a}z\u{a}van$^{1,*}$, S. Saner$^{1,*}$,  M. Minder$^{1}$, A. C. Hughes$^{1}$, R. T. Sutherland$^{2}$, D. M. Lucas$^{1}$, R. Srinivas$^{1}$, C. J. Ballance$^{1, 3}$ \\
\normalsize{$^{1}$Department of Physics, University of Oxford, Clarendon Laboratory, Parks Road, Oxford OX1 3PU, UK}\\
\normalsize{$^{2}$Department of Electrical and Computer Engineering, University of Texas at San Antonio, San Antonio, Texas 78249, USA}\\
\normalsize{$^{3}$Oxford Ionics, Oxford, OX5 1PF} \\
\normalsize{$^*$These authors contributed equally. \\
Email: oana.bazavan@physics.ox.ac.uk, sebastian.saner@physics.ox.ac.uk}}
\date{\today}

\begin{abstract}
A single bichromatic field near-resonant to a qubit transition is typically used for $\hat{\sigma}_x$ or $\hat{\sigma}_y$ M{\o}lmer-S{\o}rensen type interactions in trapped ion systems. Using this field configuration, it is also possible to synthesize a $\hat{\sigma}_z$ spin-dependent force by merely adjusting the beat-note frequency. Here, we expand on previous work and present a comprehensive theoretical and experimental investigation of this scheme with a laser near-resonant to a quadrupole transition in $^{88}$Sr$^+$. Further, we characterise its robustness to optical phase and qubit frequency offsets, and demonstrate its versatility by entangling optical, metastable, and ground state qubits.
\end{abstract}
\maketitle

Trapped ion systems are used for quantum computation~\cite{cirac1995quantum, monroe1995demonstration, Wineland1998}, quantum simulation~\cite{ blatt2012quantum}, metrology, and sensing~\cite{schmidt2005spectroscopy, wolf2016non}. These applications typically require coupling of the internal spin states of the ions to their shared motion via a spin-dependent force (SDF)~\cite{blatt2008entangled}. These SDFs can arise from the intensity gradient of applied lasers, or from magnetic field gradients~\cite{Wineland1998, mintert2001quantum, ospelkaus2008oscillating, johanning2009individual, ospelkaus2011microwave}. The basis of the SDF, the specific Pauli spin-operator that it corresponds to, depends on its particular physical implementation. For example, time-varying ac Stark shifts can be used to implement $\hat{\sigma}_z$-type interactions~\cite{milburn2000ion,Leibfried2003,ballance2016high}, while a bichromatic field near-resonant to the qubit transition can be used to implement $\hat{\sigma}_x$ or $\hat{\sigma}_y$ Mølmer-Sørensen (MS) type interactions~\cite{sorensen1999quantum, Soerensen2000, sackett2000experimental, gaebler2016high, erhard2019characterizing}. The basis of the SDF determines the fields required, the applicability of the interaction to different qubit types, and its sensitivity to errors.

The $\hat{\sigma}_z$-type interaction can be made insensitive to errors from qubit frequency offsets or dephasing, which commute with the interaction~\cite{Leibfried2003}, using spin-echo sequences~\cite{ hahn1950_spin, levitt1986composite}.
This interaction typically requires fields off-resonant from the qubit transition and is intrinsically insensitive to phase fluctuations of the driving field. Conventional implementations relying on dipole transitions are incompatible with magnetic-field insensitive ‘clock’ qubits \cite{lee2005phase}, desirable for their long coherence times~\cite{langer2005long,wang2021single}. Nevertheless, recent experiments have used quadrupole transitions with ground state qubits~\cite{aolita2007high, baldwin2021high} or dipole transitions with optical qubits~\cite{sawyer2021wavelength, clark2021high} that enable the interaction to be applied to clock qubits. On the other hand, the MS-type interaction requires near-qubit frequency fields. This interaction is readily applicable to clock qubits, but is sensitive to the phase of the driving field~\cite{lee2005phase, haljan2005spin}. Additionally, the same fields that drive the MS interaction can be tuned to perform single-qubit rotations.

Using near-qubit frequency fields to implement a $\hat{\sigma}_z$-type interaction minimises the number of required fields and improves its robustness to errors commuting with the $z$-basis.
Moreover, it enables a wider range of interactions, relevant for applications such as quantum simulation \cite{ Britton2012, bermudez2017long}. Another important consideration is the applicability of a given scheme to a variety of qubit encodings, for example in the ‘omg’-type architecture which requires control of optical, metastable, and ground state qubits~\cite{yang2021realizing, allcock2021omg}. Finally, the wavelengths of the required fields also form an important practical criterion. For instance, the wavelengths of quadrupole transitions are usually in the red, which are more favorable for current integrated optics technologies \cite{niffenegger2020integrated, Mehta2020} and to reduce trap charging effects \cite{wang2011laser}.

In this work, we investigate a technique for implementing a laser-based $\hat{\sigma}_z$ SDF using a bichromatic field on a quadrupole transition, proposed in Ref.~\cite{Roos_2008} and inspired by recent work in Refs.~\cite{sutherland2019versatile,srinivas2021high} for laser-free interactions. This SDF-mechanism was proposed and previously demonstrated using low SDF amplitudes in Refs.~\cite{kim2007geometric,monz2009realization, gorman2018engineering}. We present a comprehensive theory treatment of the interaction as well as in-depth experimental characterisation of the SDF for both low and high SDF amplitudes. Moreover, we demonstrate the versatility of the SDF created by using it to entangle optical, metastable, and ground state qubits.

\begin{figure}
    \includegraphics[width=1.0\linewidth]{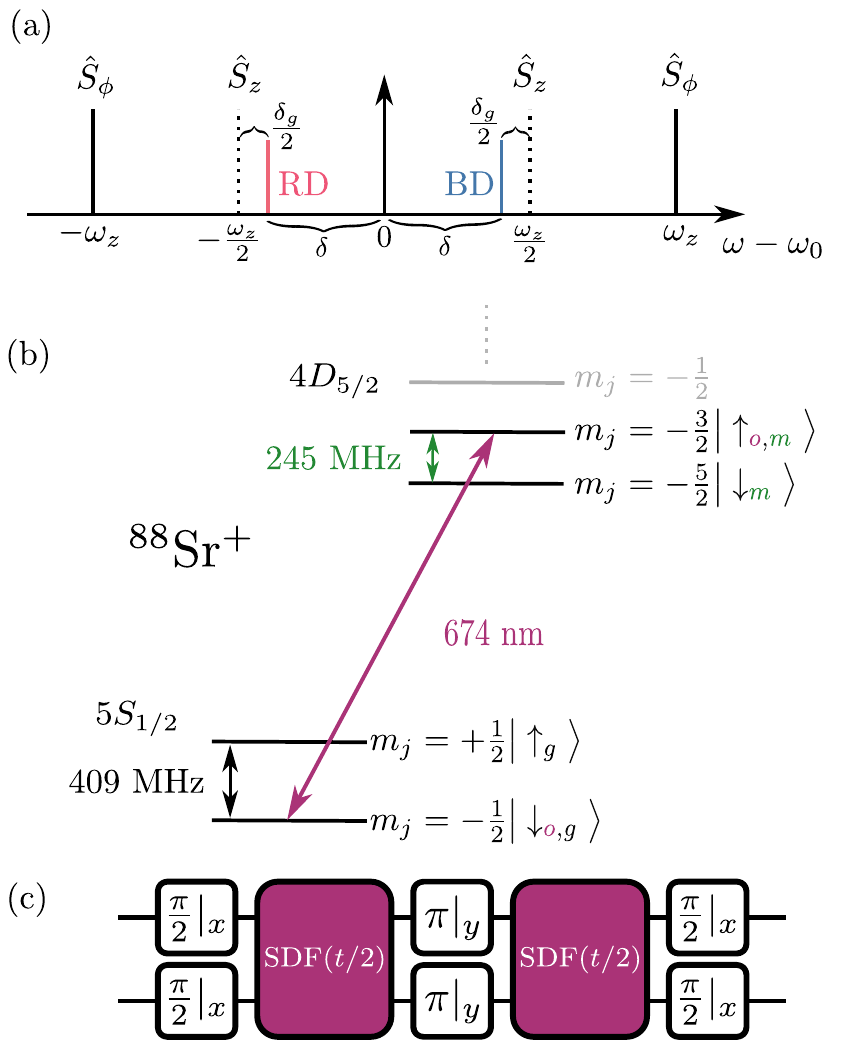}
    \caption{(a) Frequency configuration of the bichromatic laser field. We apply two 674\,nm tones (RD and BD) that are symmetrically detuned from the optical qubit transition frequency, $\omega_0$, by $\delta$. To implement the $\hat{S}_z$ interaction, we set $\delta = (\omega_z -\delta_g)/2$, where $\delta_g$ corresponds to the detuning of the effective spin-dependent interaction, and $\omega_z$ is the motional mode frequency. (b) Overview of the relevant atomic structure for $^{88}$Sr$^+$. Using the same bichromatic field on the 674\,nm transition, we implement this scheme on the optical, ($\ket{\downarrow_o}\leftrightarrow\ket{\uparrow_o}$), metastable ($\ket{\downarrow_m}\leftrightarrow\ket{\uparrow_m}$), and ground state ($\ket{\downarrow_g}\leftrightarrow\ket{\uparrow_g}$) qubits. (c) Pulse sequence for entangling operations. The single-qubit operations ($\pi$-pulses and $\pi/2$-pulses) are performed with a 674\,nm laser for the optical qubit and with a radiofrequency field from an antenna for the ground and metastable state qubits. The phase of the $\pi$-pulse is orthogonal to that of the $\pi/2$-pulses.}
    \label{Figure1}
\end{figure}

To understand this technique, let us first consider a collection of $n$ spins coupled to a motional mode by a bichromatic field~\cite{Roos_2008}. The field is composed of two tones which are symmetrically detuned from the optical qubit frequency, $\omega_0$, by $\delta$, as shown in Fig.~\ref{Figure1}(a). These fields give rise to an interaction~\cite{Soerensen2000}
\begin{equation}
    \begin{split}
    \hat{H} =&\ \hbar \Omega \hat{S}_{\phi-\pi/2} \cos{(\delta t)} \\
    &+\hbar\Omega \eta \hat{S}_{\phi} \cos{(\delta t)} (\hat{a}e^{-i \omega_z t} + \hat{a}^\dagger e^{i \omega_z t}),
    \end{split}
\label{eq:H_MS}
\end{equation}

\noindent where $\Omega$ denotes the Rabi frequency for each tone and $\eta$ the Lamb Dicke factor~\cite{Wineland1998}. The spin operator for $n$ ions is defined as $\hat{S}_{\phi} = \sum_{i=1}^{n}\hat{\sigma}_{\phi}^{(i)}$ with $\hat{\sigma}_{\phi}^{(i)} = \cos\phi \hat{\sigma}_x^{(i)} + \sin\phi \hat{\sigma}_y^{(i)}$, \footnote{$\hat{\sigma}_{x,y}^{(i)} =\underbrace{\hat{\mathbb{I}}_2 \otimes ...\otimes \hat{\mathbb{I}}_2}_{i-1} \otimes \hat{\sigma}_{x,y}\otimes \underbrace{\hat{\mathbb{I}}_2 \otimes ...\otimes \hat{\mathbb{I}}_2}_{n-i}$} and $\hat{a}^\dagger$($\hat{a}$) denotes the creation (annihilation) operator of the motional mode. The phase $\phi = (\phi_{\mathrm{BD}} + \phi_{\mathrm{RD}})/2$ is the mean optical phase between the red (RD) and the blue (BD) detuned tones. The above expression is in the interaction picture with respect to the qubit frequency $\omega_0$, and the motional mode frequency $\omega_z$, where 
we applied the rotating wave approximation
with respect to $\omega_0$. The first term drives spin flips (carrier term) and the
second term couples the spins to the motional mode (sideband term). Crucially, these two terms do not commute. The resulting dynamics can be simplified by moving into the interaction picture with respect to the carrier term, details are in Appendix~\ref{appendix:theory}. Following the derivation in Refs.~\cite{Roos_2008, sutherland2019versatile}, we obtain

\begin{equation}\label{eq:H_full}
    \begin{split}
        \hat{H}_I =&\ \hbar \eta \Omega (\hat{a} e^{-i \omega_z t} + \hat{a}^\dagger e^{i \omega_z t}) \times \\
        &\Big[ \hat{S}_{\phi}\sum_{n=0}^\infty \big(J_{2n}(2\Omega/\delta)+J_{2n+2}(2\Omega/\delta)\big)\cos{((2n+1)\delta t )}\\
        &-\hat{S}_z\sum_{n=1}^\infty \big(J_{2n-1}(2\Omega/\delta)+ J_{2n+1}(2\Omega/\delta)\big) \sin{(2n\delta t)}\Big],
    \end{split}
\end{equation}
\noindent where $\hat{S}_z = \sum_{i=1}^{n} \hat{\sigma}_{z}^{(i)}$ and $J_n$ are the Bessel functions of the first kind. Equation~\ref{eq:H_full} reveals an infinite series of resonances which can be selectively driven. In contrast to Ref.~\cite{sutherland2019versatile}, here both the carrier and the sideband term oscillate at the same frequency $\delta$, as they originate from the same source. Different choices of $\delta$ will drive different interactions; when $\delta\approx\omega_z/(2n+1)$ we drive MS-type interactions, while $\delta\approx\omega_z/(2n)$ corresponds to $\hat{S}_z$-type interactions. For ${\delta \approx \omega_z}$, we obtain the conventional MS interaction with coupling strength modulated by $J_0 + J_2$ and a spin-dependent force in the $\hat{S}_{\phi}$~basis. However, by choosing $\delta \approx \omega_z/2$, the near-resonant term
\begin{equation}\label{eqn:H_I}
    \begin{split}
            \hat{H}_I =& -\hbar \eta\Omega (J_1(2\Omega/\delta)+ J_3(2\Omega/\delta))\times \\
            &\sin{(2\delta t)} \hat{S}_z(\hat{a} e^{-i \omega_z t} + \hat{a}^\dagger e^{i \omega_z t}),
    \end{split}
\end{equation}
drives an $\hat{S}_z$ interaction~\cite{Roos_2008} with a coupling strength modulated by $J_1 + J_3$. The effective coupling strength is then ${\Omega_{\mathrm{eff}}=\eta \Omega (J_1(2\Omega/\delta)+ J_3(2\Omega/\delta))}$. Importantly, this interaction is insensitive to the mean optical phase $\phi$ of the bichromatic laser field, in contrast to the MS-interaction $\hat{S}_{\phi}$.
This SDF can be employed for creating entanglement between qubits using standard techniques. Moreover, as this effective $\hat{S}_z$ force is derived from an interaction that couples directly to the qubit levels, it could also be used with field-insensitive `clock' qubits. In the regime where ${\Omega \ll \delta}$, Eq.~\ref{eqn:H_I} reproduces the effective Hamiltonian described in Ref.~\cite{kim2007geometric}, which was experimentally demonstrated in Ref.~\cite{monz2009realization, gorman2018engineering}.

We experimentally demonstrate this interaction on trapped $^{88}$Sr$^+$ ions in a 3D radiofrequency Paul trap~\cite{schafer2018fast,thirumalai2019high}. We use a single 674\,nm beam with two tones symmetrically detuned from the \(\ket{5S_{1/2},\,m_j = -\frac{1}{2}} \leftrightarrow \ket{4D_{5/2},\,m_j = -\frac{3}{2}}\) quadrupole transition at a 146\,G magnetic field. For both the single- and two-ion experiments, we choose to drive the axial in-phase mode with $\omega_z/2\pi \approx$ 1.2~MHz. We obtain carrier Rabi frequencies of $\Omega/2\pi = 0.14 -1.17$~MHz using a beam waist radius of 20\,$\mu$m and laser powers between 0.5 and 35\,mW. To generate the SDF acting in the $\hat{S}_z$ basis, we set $\delta = (\omega_z-\delta_g)/2$ as shown in Fig.~\ref{Figure1}(a), where $\delta_g$ is the detuning of the $\hat{\sigma}_z$ interaction from resonance. Unless stated otherwise, the pulses generating the $\hat{S}_z$ interaction are embedded in a spin-echo Ramsey sequence~\cite{hahn1950_spin, levitt1986composite} (see Fig.~\ref{Figure1}(c)). We set the phase of the SDF in the second pulse to match the phase of the SDF at the beginning of the first pulse relative to the ion motion. To transition smoothly into the bichromatic interaction picture, we ramp the bichromatic field on and off  with a sin$^2$ ramp shape. The ramp duration should be long compared to $1/\delta$, which is approximately 2\,$\mu$s. We use ramp durations of 5 or 10\,$\mu$s.

\begin{figure}
  \centering
    \includegraphics[width=1.\linewidth]{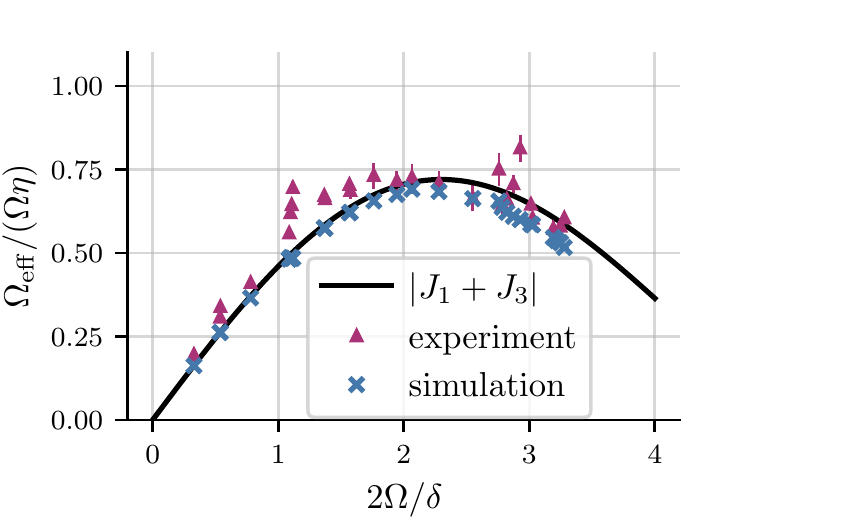}
    \caption{Spin-dependent force strength, $\Omega_{\mathrm{eff}}$, normalised by $\Omega\eta$ (see text), versus $2\Omega/\delta$, as measured on a single ion, with $\delta\approx\omega_z/2$. We show the experimental data (magenta triangles) extracted from measuring the spin-dependent force amplitude by applying the interaction for variable durations and fitting the dynamics~\ref{appendix:sdf_extraction}. The error bars are inferred from the fits, showing the 68\% confidence intervals. The data agree with the theory, $|J_1(2\Omega/\delta) +J_3(2\Omega/\delta)|$ (black line). 
    To validate the approximations in Eq.\,\ref{eqn:H_I} we perform numerical simulations of Eq.\,\ref{eq:H_MS} (blue crosses), using the experimental parameters, and we follow the same extraction procedure for $\Omega_{\mathrm{eff}}$ as for the measured data.
    }
    \label{Figure2}
\end{figure}

We experimentally verify two important characteristics of the SDF using the optical qubit states ${\ket{\downarrow_o}\equiv\ket{5S_{1/2},\,m_j = -\frac{1}{2}}}$ and  ${\ket{\uparrow_o}\equiv\ket{4D_{5/2},\,m_j = -\frac{3}{2}}}$, on a single ion. Firstly, we show that the SDF magnitude follows the predicted Bessel function dependence, $J_1(2\Omega/\delta) + J_3(2\Omega/\delta)$ (see Fig.~\ref{Figure2}).
For each data point, we apply the sequence shown in Fig.~\ref{Figure1}(c) to a single qubit and measure the single-ion population as a function of the SDF pulse duration $t$.
We fit these dynamics to an effective SDF model with magnitude, $\Omega_{\mathrm{eff}}$, normalised by $\Omega$ and the Lamb Dicke factor, $\eta\approx0.054$, which are independently determined~\ref{appendix:sdf_extraction}. We repeat the experiment for different values of $2\Omega/\delta$ by changing both the Rabi frequency $\Omega$ and detuning $\delta$. We also perform numerical simulations of Eq.\,\ref{eq:H_MS}, assuming the experimental parameters and applying the same method for extracting $\Omega_{\mathrm{eff}}$. For larger values of $2\Omega/\delta$ the numerical simulation deviates from the theory curve, which does not include any dynamics during the ramp.

\begin{figure}
  \centering
    \includegraphics[width=1.\linewidth]{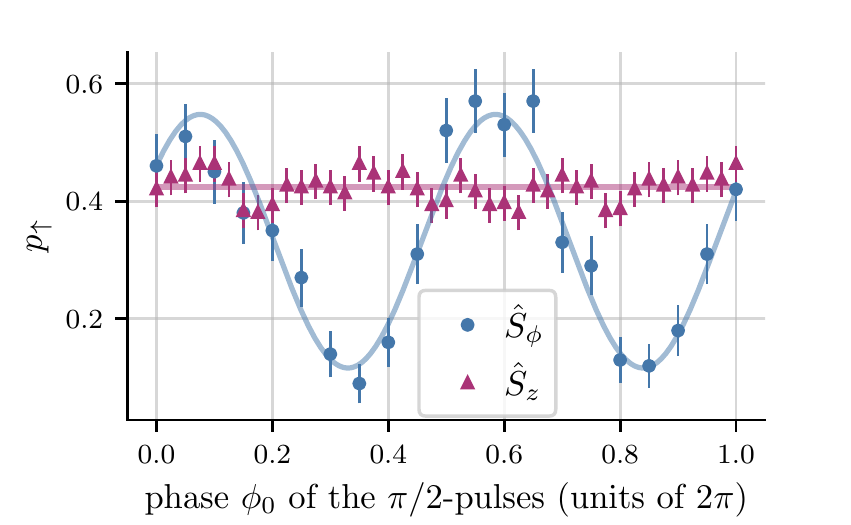}
    \caption{Verification of the spin-dependent force (SDF) basis. We scan the phase $\phi_0$ of the $\pi/2$-pulses relative to the SDF. We map the spin-dependent displacement generated by both $\hat{S}_z$ (magenta triangles) and  $\hat{S}_\phi$ (blue circles) forces onto $\ket{\uparrow_o}$ and measure the population $p_\uparrow$ (see text). As expected, the signal for the $\hat{S}_z$ SDF has no dependence on $\phi_0$, while the $\hat{S}_\phi$ SDF does. The solid lines are a guide to the eye of the expected dependence. Error bars indicate 68\% confidence intervals.}
    \label{Figure 3}
\end{figure}

Secondly, we verify that the SDF acts in the $\hat{S}_z$-basis and compare it to the MS case. We embed the SDF in a sequence similar to Fig.~\ref{Figure1}(c), applied to a single qubit, but omit
the $\pi$-pulse, and the second SDF pulse. Then, we scan the phase $\phi_0$ of both the $\pi/2$-pulses relative to the SDF drive (Fig.~\ref{Figure 3}). Changing $\phi_0$ modifies the initial superposition of the spin states; the SDF acting on its eigenstate will effect no spin-dependent displacement. We map the generated spin-dependent displacement onto the spin state and measure the population $p_{\uparrow}$ in $\ket{\uparrow_o}$. We set the detuning and duration of the SDF to generate a large enough spin-dependent displacement such that $p_\uparrow\approx0.5$. For this scheme, $p_\uparrow$ is independent of $\phi_0$. Hence, the SDF basis is orthogonal to the $\hat{S}_x -\hat{S}_y$ plane and corresponds to $\hat{S}_z$. In contrast, for the MS-type interaction, which acts in the $\hat{S}_{\phi}$ basis, the spin-dependent displacement depends on whether the superposition state after the first $\pi/2$-pulse is aligned or orthogonal to the SDF basis. 
In these experiments, we switch the basis of the force by setting the detuning to $\delta\approx\omega_z/2$ or $\delta\approx\omega_z$ for the $\hat{S}_z$ or $\hat{S}_\phi$ interactions, respectively.

Another important aspect of this $\hat{S}_z$ force is its suitability to any qubit encoding where one or both qubit levels are part of the quadrupole transition used to generate the SDF, as these states are all eigenstates of the interaction.
This property does not hold for the $\hat{S}_\phi$ case. We demonstrate the versatility of this SDF by implementing a geometric-phase two-qubit $\hat{S_z}$ entangling gate on the optical, metastable, and ground state qubits following the pulse sequence in Fig.~\ref{Figure1}(c). The experimental parameters for $\Omega$ and $\delta$ for all three demonstrations correspond to $2\Omega/\delta \approx 1.6$, the argument of the Bessel function in Fig.\,\ref{Figure2}. The quoted gate durations correspond to $4\pi/\delta_g$. The total duration of the SDF pulses includes an additional 20$\,\mu$s for the ramps. For each of the qubit types, we create the entangled state $\frac{1}{\sqrt{2}}(\ket{\downarrow\downarrow} + \ket{\uparrow\uparrow})$. We infer the fidelity of the created Bell state by measuring the populations and the parity, $1-2(p_{\uparrow\downarrow} + p_{\downarrow\uparrow})$. For the parity measurement, we add a $\pi/2$ analysis pulse with a variable phase~\cite{sackett2000experimental} after the pulse sequence in Fig.~\ref{Figure1}(c). We report the fidelity $\mathcal{F}$ without correcting for any state preparation and measurement (SPAM) error $\bar{\epsilon}$, which we report separately~\cite{ballance2017high}.

\begin{figure}
    \includegraphics{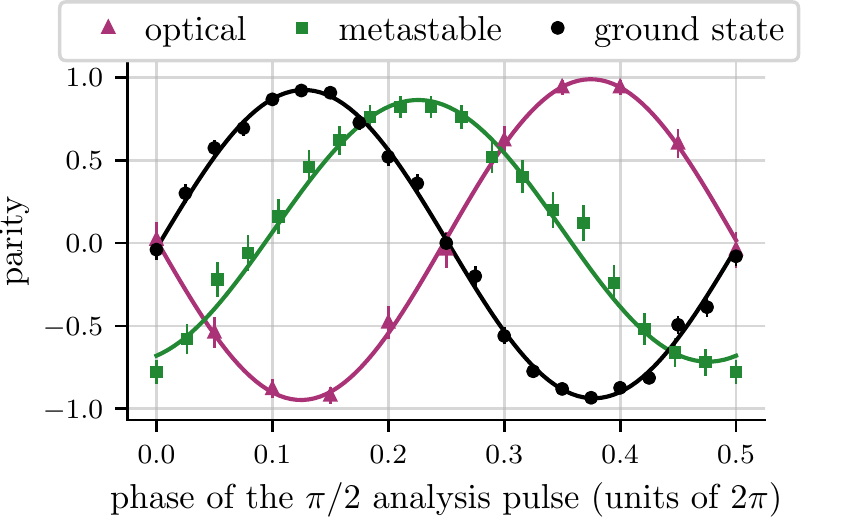}
    \caption{Example parity scans for two-qubit entangled states created in the optical (magenta triangles), metastable (green squares) and ground (black circles) state qubits. We scan the $\pi/2$ analysis pulse phase and measure the parity (see text). Error bars indicate 68\% confidence intervals, and lines are fits to the data.}
    \label{fig:parity_scans}
\end{figure}

\textit{Optical qubit}: We start with the optical qubit, where the SDF couples to both qubit states $\ket{\downarrow_o}$ and $\ket{\uparrow_o}$. We obtain a Bell-state fidelity of $\mathcal{F}= 0.930(3)$ for a gate duration of $\approx$70 $\mu$s at 9.1\,mW. The SPAM error is $\bar{\epsilon}=0.0016(2)$. An example parity scan is shown in Fig.~\ref{fig:parity_scans}. In contrast to MS-gate implementations, the phase of the entangled state is fixed and does not depend on the average phase of the two bichromatic field tones.
The SDF pulses and the single-qubit rotations were implemented using the same 674\,nm laser.

\begin{figure}[!t]
  \centering
    \includegraphics[width=\linewidth]{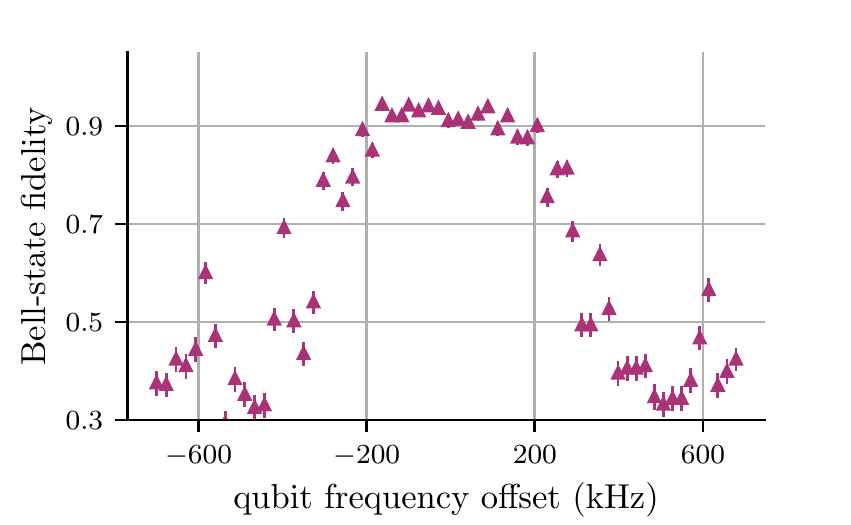}
    \caption{Bell-state fidelity $\mathcal{F}$ versus the applied qubit frequency offset for the optical qubit. The Bell-state fidelity remains unchanged above $0.9$ over a range of $\pm$150\,kHz. Fidelity values below 0.5 indicate a change in the phase of the created Bell state.}
    \label{fig:fidelity_vs_det_and_phase}
\end{figure}

To demonstrate the robustness of the gate mechanism to $\hat{\sigma}_z$-type errors, we measure the Bell-state fidelity as a function of an applied qubit frequency offset, where we shift both tones by the same offset, during the gate pulses (Fig.~\ref{fig:fidelity_vs_det_and_phase}). From $-150$ to $150$\,kHz, the Bell-state fidelity values are statistically consistent with each other and stay above $0.9$. 

\textit{Metastable qubit}: We next entangle the metastable qubit states ${\ket{\downarrow_m}\equiv \ket{4D_{5/2},\,m_j = -\frac{5}{2}}}$ and ${\ket{\uparrow_m}\equiv \ket{4D_{5/2},\,m_j = -\frac{3}{2}}}$, split by $\approx$245\,MHz (see Fig.~\ref{Figure1}(b)). We obtain a Bell-state fidelity of ${\mathcal{F} = 0.859(5)}$ for a gate duration of $\approx$140~$\mu$s at $9.5$~mW laser power. The SPAM error is $\bar{\epsilon}=0.0164(7)$. As the SDF only couples to the $\ket{\uparrow_m}$ state, the differential force between the qubit states is half that of the optical qubit. Hence, the gate duration is about twice as long as for the optical qubit case using similar laser power. 
The single-qubit rotations for the metastable qubit are performed with a radiofrequency (rf) antenna; crucially we require no defined phase relationship between the rf source and the 674\,nm laser for the SDF.

As the Zeeman splitting between neighbouring $4D_{5/2}$ states is almost identical, the qubit states need to be isolated to prevent population leakage during the single-qubit rotations. To break the degeneracy, we light-shift the $\ket{\uparrow_m}$ level by applying a 674\,nm beam blue detuned from the optical quadrupole transition ($\ket{\downarrow_o} \leftrightarrow \ket{\uparrow_o}= \ket{\uparrow_m}$) during the single-qubit rotations. Using a detuning of 2.8\,MHz and power of 20\,mW, we shift the qubit transition by $\approx145$\,kHz from the neighbouring transition. We attribute most of the additional infidelity compared to the optical qubit case to imperfect state initialisation and single-qubit rotations. We show a sample parity scan for the metastable qubit in Fig.~\ref{fig:parity_scans}. The detuned 674\,nm field used during the single-qubit operations also induces a phase shift on the entangled state which results in a phase offset in the parity scan. 

\textit{Ground state qubit}: Finally, we demonstrate a two-qubit entangling gate using the ground state qubits ${\ket{\downarrow_g}\equiv \ket{5S_{1/2},\,m_j = -\frac{1}{2}}}$ and ${\ket{\uparrow_g}\equiv\ket{5S_{1/2},\,m_j = \frac{1}{2}}}$, split by 409\,MHz (see Fig.~\ref{Figure1}(b)). We obtain a Bell-state fidelity of $\mathcal{F} = 0.949(4)$ for a gate duration of $\approx$140~$\mu$s at 9~mW laser power. The SPAM error is $\bar{\epsilon}=0.0036(4)$. Similarly, the single-qubit rotations are implemented via the rf antenna. Again, the gate duration is doubled compared to the optical qubit case as the SDF only couples to $\ket{\downarrow_g}$. 

We have shown that the technique is applicable to three different qubit types using the exact same bichromatic field. Within the context of omg architectures, this simplifies the technical requirements as it enables the manipulation of all three qubit encodings using a single quadrupole laser. However, there might be additional considerations for multi-ion chains. For example, a global laser beam would couple ions whose qubits are either in the ground or metastable manifold. Single-ion addressing of the gate drive preserves orthogonality in qubit sub-space. However, most omg-applications would require single-ion addressing which would mitigate this issue. In addition, auxiliary states in either manifold that do not couple to the bichromatic field can be used to shelve quantum states temporarily. Another application of this scheme, not explored in this paper, is entangling ions that use different qubits e.g.~metastable-ground state, optical-ground state.

In terms of the errors associated with the entangling operations using this SDF scheme, we are able to achieve Bell-state fidelities comparable to those achieved by the MS-interaction in the same system. We believe the main sources of infidelity are phase noise from the 674\,nm laser and excitation of other motional mode resonances. The phase of the entangled state, and hence the fidelity, is insensitive to slow drifts in the optical phase for the $\hat{S}_z$ interaction. However, similar to the MS interaction, it is still sensitive to phase noise that is fast compared to the duration of an entangling gate. Additionally, single-qubit rotations on the optical qubit can also induce errors. We observed a loss in contrast from the spin-echo Ramsey sequences, Fig.~\ref{Figure1}(c) without any SDF pulses, at delays close to the durations used for the entangling gates. The contrast loss on the two-ion population $p_{\uparrow\downarrow}+p_{\downarrow\uparrow}$ was $0.035(3)$ for the optical qubit encoding.
This indicates the presence of coloured phase noise on the 674\,nm laser which could significantly affect the gate dynamics. For the ground state and the metastable qubit, the contrast loss was dominated by SPAM errors. Furthermore, there are additional resonances close to $\delta\approx\omega_z/2$ as the laser couples to all the motional modes of the ions~\ref{appenxix:spectator_modes}. We observed that sideband cooling of the additional motional modes corresponding to these nearby resonances mitigates this error contribution. 

In conclusion, we present an in-depth investigation of a spin-dependent force in the $\hat{\sigma}_z$-basis created using a bichromatic laser-field near-resonant with a quadrupole transition. By only changing the beat-note frequency of the near-qubit tones, we can toggle the interaction basis from the typical $\hat{\sigma}_\phi$ to $\hat{\sigma}_z$.
We characterise this interaction and investigate its robustness to $\hat{\sigma}_z$-type errors, such as qubit frequency offsets, and show that the basis of the SDF does not depend on the optical phase of the driving field. This SDF is well suited to manipulate a variety of qubit states. We use the same laser configuration to entangle optical, metastable, and ground state qubits, which is important for recently proposed trapped-ion architectures~\cite{yang2021realizing, allcock2021omg}. While demonstrated for field-sensitive transitions (there are no clock qubits in $^{88}\mathrm{Sr}^+$), this method could be extended to field-insensitive clock qubits. Moving beyond quadrupole transitions, this method can also be used with two-photon Raman and magnetic dipole transitions in other trapped ion systems. 

We thank David Allcock, Gabriel Araneda, and Alejandro Bermúdez for comments on the manuscript. This work was supported by the UK EPSRC Hub in Quantum Computing and Simulation (EP/T001062/1) and the US Army Research Office (W911NF-20-1-0038). 

\appendix
\renewcommand{\thefigure}{B.\arabic{figure}}
\setcounter{figure}{0}


\section{Derivation of $\hat{\sigma}_z$ interaction}\label{appendix:theory}
In this section, we present a more general form of Eq.\,\ref{eq:H_MS} and show an intermediate step in the derivation of Eq.~\ref{eq:H_full}. As in the main text, we start with the Hamiltonian that describes the coupling of a bichromatic field with a beat-note frequency $2\delta$ and $n$ spins
\begin{equation}
    \begin{split}
        \hat{H} =&\  \hbar \Omega \hat{S}_{\phi-\pi/2} \cos{(\delta t -\zeta)} \\
        &+ \hbar\Omega \eta \hat{S}_{\phi} \cos{(\delta t-\zeta)} (\hat{a}e^{-i \omega_z t} + \hat{a}^\dagger e^{i \omega_z t}),
    \end{split}
\end{equation}
where $\Omega$ denotes the Rabi frequency for each tone and $\eta$ the Lamb Dicke factor~\cite{Wineland1998}. The spin operator for $n$ ions is defined as $\hat{S}_{\phi} = \sum_{i=1}^{n}\hat{\sigma}_{\phi}^{(i)}$ with $\hat{\sigma}_{\phi}^{(i)} = \cos\phi \hat{\sigma}_x^{(i)} + \sin\phi \hat{\sigma}_y^{(i)}$, and $\hat{a}^\dagger$($\hat{a}$) denotes the creation (annihilation) operator of the motional mode. The phase $\phi = (\phi_{\mathrm{BD}} + \phi_{\mathrm{RD}})/2$ is the mean optical phase between the red (RD) and the blue (BD) detuned tones. Here, we also include the phase $\zeta = (\phi_{\textrm{BD}} -\phi_{\textrm{RD}})/2$, the difference in optical phase between the two tones. In our field configuration, we have control over $\zeta$ while $\phi$ is drifting. The above expression is in the interaction picture with respect to the qubit frequency $\omega_0$ and the motional mode frequency $\omega_z$, where we applied the rotating wave approximation with respect to $\omega_0$.
We express the Hamiltonian in the interaction picture with respect to the carrier coupling $\hbar \Omega \hat{S}_{\phi-\pi/2} \cos{(\delta t - \zeta)}$ and we use $[\hat{S}_{\phi-\pi/2}, \hat{S}_\phi]= i\hat{S}_z$ to obtain
\begin{equation}\label{eq:H_full_appendix}
\begin{split}
    \hat{H}_I =&\ \hbar \eta \Omega \cos{(\delta t-\zeta)}(\hat{a} e^{-i \omega_z t} + \hat{a}^\dagger e^{i \omega_z t}) \times\\
    &\Big[\hat{S}_{\phi}\big(J_0(2\Omega/\delta) + 2\sum_{n=1}^\infty J_{2n}(2\Omega/\delta)\cos{(2n(\delta t -\zeta))}\big) \\
    &- 2 \hat{S}_z \sum_{n=1}^\infty J_{2n-1}(2\Omega/\delta)\sin{((2n-1)(\delta t -\zeta))}\Big],
\end{split}
\end{equation}
as shown in Refs.~\cite{Roos_2008,sutherland2019versatile}. Simplifying Eq.\,\ref{eq:H_full_appendix} further, the Hamiltonian is
\begin{equation}
    \begin{split}
        \hat{H}_I =&\ \hbar \eta \Omega (\hat{a} e^{-i \omega_z t} + \hat{a}^\dagger e^{i \omega_z t}) \times \\
        &\Big[ \hat{S}_{\phi}\sum_{n=0}^\infty \big(J_{2n}(2\Omega/\delta)+J_{2n+2}(2\Omega/\delta)\big)\cos{((2n+1)(\delta t-\zeta))}\\
        &-\hat{S}_z\sum_{n=1}^\infty \big(J_{2n-1}(2\Omega/\delta)+ J_{2n+1}(2\Omega/\delta)\big) \sin{(2n(\delta t-\zeta))}\Big].
    \end{split}
\end{equation}
By setting $\delta \approx \omega_z/2$, the leading term is
\begin{equation}
    \begin{split}
        \hat{H}_I \approx -&\hbar \eta\Omega (J_1(2\Omega/\delta)+ J_3(2\Omega/\delta))\times\\
        &\sin{(2(\delta t -\zeta))} \hat{S}_z(\hat{a} e^{-i \omega_z t} + \hat{a}^\dagger e^{i \omega_z t}),
    \end{split}
\end{equation}
which corresponds to a $\hat{\sigma}_z$ SDF~\cite{Roos_2008}. In the main text, we set $\zeta=0$ for simplicity. In practice, we adjust $\zeta$ such that the phase of the SDF for the second pulse matches the phase of the SDF of the first pulse relative to the ion motion.

\section{Experimental Details}
\subsection{Spin-dependent force extraction}\label{appendix:sdf_extraction}

\begin{figure}[ht]
  \centering
    \includegraphics[]{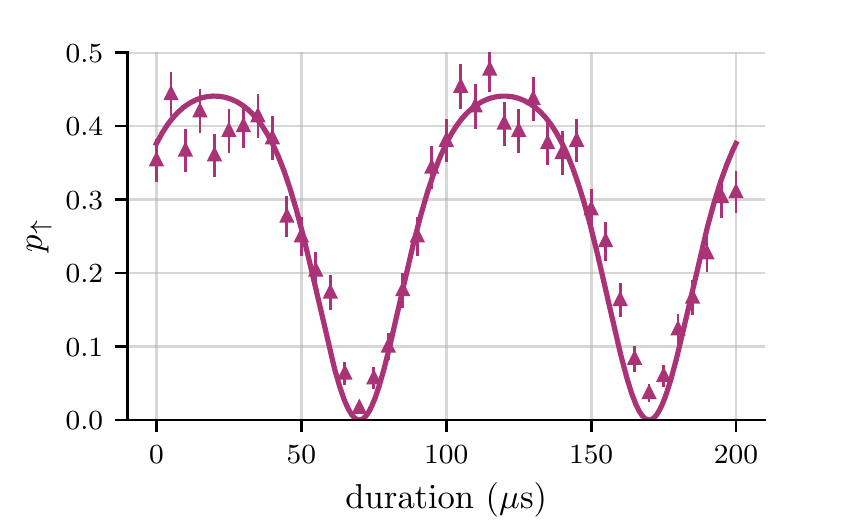}
    \caption{Example dynamics of the SDF applied to a single ion in the optical qubit. We plot the population of the ion in the $\ket{\uparrow}$ as a function of the pulse duration. For this scan, $2\Omega/\delta \approx 1.1$. Error bars indicate 68\% confidence intervals, and the line is a fit to the data following Eq.~\ref{eq:effective_force}.}
    \label{fig:sdf_dynamics_example}
\end{figure}

For each data point in Fig.~\ref{Figure2}, we apply the SDF to a single ion and measure the population $p_\uparrow$ as a function of the SDF duration, as shown in Fig.~\ref{fig:sdf_dynamics_example}. We fit these dynamics to obtain the magnitude of the SDF, $\Omega_{\mathrm{eff}}$. We use an effective SDF model~\cite{haljan2005spin}, which is described by

\begin{align}
    \hat{H}_{\mathrm{eff}} &= \hbar \Omega(t)(\hat{a} e^{-i \delta_g t} + \hat{a}^\dagger e^{i \delta_g t})\hat{S}_z,
    \label{eq:effective_force}
\end{align}
where the Rabi frequency of the interaction $\Omega(t)$ follows
\begin{align}
 \Omega(t) &=
 \begin{cases}
 \Omega_{\mathrm{eff}} \sin{\left(\frac{\pi}{2} \frac{t}{t_{\mathrm{ramp}}}\right)}^2, & t<t_{\mathrm{ramp}}\\
 \Omega_{\mathrm{eff}}, & t_{\mathrm{ramp}} \le t \le t_{\mathrm{total}}-t_{\mathrm{ramp}}\\
 \Omega_{\mathrm{eff}} \sin{\left(\frac{\pi}{2} \frac{t_{\mathrm{total}}-t}{t_{\mathrm{ramp}}}\right)}^2, & t >t_{\mathrm{total}}-t_{\mathrm{ramp}},\\
 \end{cases}
\end{align}
where $t_{\mathrm{total}}$ is the total duration of a single SDF pulse including the ramps. The duration over which the SDF is turned on and off is described by $t_{\mathrm{ramp}}$. The SDF described in Eq.~\ref{eq:effective_force} gives rise to a spin-dependent displacement ${\hat{U}(t) = \hat{D}(\alpha(t))\ket{\uparrow}\bra{\uparrow}+\hat{D}(-\alpha(t))\ket{\downarrow}\bra{\downarrow}}$, where ${\alpha(t)\propto\Omega(t)(1-e^{-\delta_gt})}$. We integrate $\alpha(t)$, keeping $t_{\mathrm{ramp}}$ and $\delta_g$ fixed. For the fit, we also take into consideration the initial occupation of the motional mode. We assume an initial thermal occupation of $\bar{n}\approx0.1$, measured through sideband asymmetry. In our 3D Paul trap, the heating rate for this motional mode is $\dot{\bar{n}}\approx35$\,quanta/s. We do not include heating in the fitting model as it has a negligible effect on the dynamics.

\begin{figure}[h!]
  \centering
    \includegraphics[]{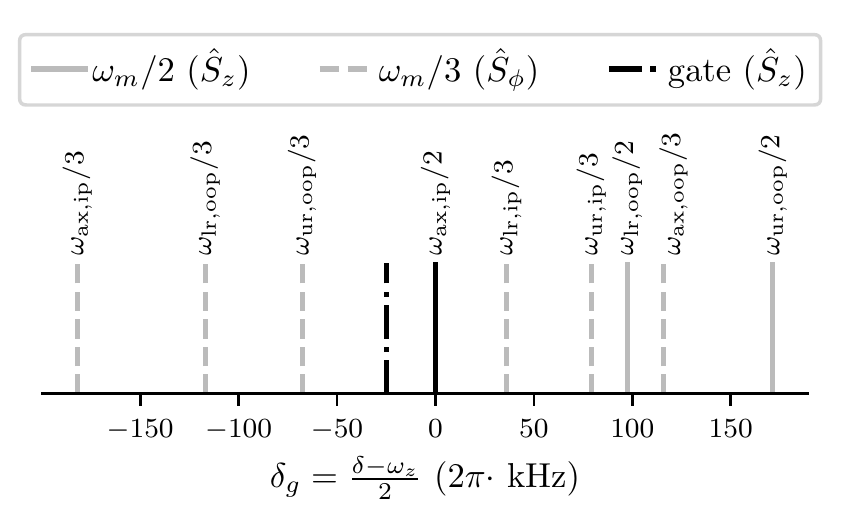}
    \caption{Spectrum of spectator resonances.
    The two-qubit $\hat{S}_z$ gate is performed on the axial in-phase mode ($\omega_z=\omega_{\mathrm{ax,ip}}$).
    The interaction resonance frequency and the detuning used for the gate are indicated by the solid black line and the dash-dotted black line, respectively.
    The other resonances are annotated with the motional mode abbreviation that they correspond to: axial out-of-phase ($\omega_{\mathrm{ax,oop}}$), lower radial in-phase ($\omega_{\mathrm{lr,ip}}$), upper radial in-phase ($\omega_{\mathrm{ur,ip}}$), lower radial out-of-phase ($\omega_{\mathrm{lr,oop}}$), upper radial out-of-phase ($\omega_{\mathrm{ur,oop}}$). Solid grey lines denote $\hat{S}_z$ couplings ($\delta \approx \omega_m/2)$ and dashed grey lines denote $\hat{S}_\phi$ couplings ($\delta\approx\omega_m/3$). }
    \label{fig:resonance_spectrum}
\end{figure}

\subsection{Spectator modes} \label{appenxix:spectator_modes}
In our system, the 674\,nm laser has a projection on all three trap axes. Hence for a two-ion crystal, the laser couples to all six motional modes. The motional modes are the axial in-phase ($\omega_{\mathrm{ax,ip}}$), axial out-of-phase ($\omega_{\mathrm{ax,oop}}$), lower radial in-phase ($\omega_{\mathrm{lr,ip}}$), upper radial in-phase ($\omega_{\mathrm{ur,ip}}$), lower radial out-of-phase ($\omega_{\mathrm{lr,oop}}$), upper radial out-of-phase ($\omega_{\mathrm{ur,oop}}$). Their frequencies are $\{{\omega_{\mathrm{ax,ip}}}, \omega_{\mathrm{ax,oop}}, \omega_{\mathrm{lr,ip}}, \omega_{\mathrm{ur,ip}}, \omega_{\mathrm{lr,oop}},
\omega_{\mathrm{ur,oop}}\}/2\pi=\{1.2, 2.0, 1.7, 1.9, 1.3, 1.4\}$\,MHz. The gate was performed on the axial in-phase mode ($\omega_z=\omega_{\mathrm{ax,ip}}$). All mode resonances in the vicinity of $\delta$ will be spectators of the interaction. The dynamics are governed by Eq.~\ref{eq:H_full} with the respective mode frequency and Lamb Dicke parameter.
Any residual population $p_{\uparrow\downarrow}+p_{\downarrow\uparrow}$ at the end of the gate sequence will result in an error. By sideband cooling the lower radial in-phase, upper radial out-of-phase, axial out-of-phase, and axial in-phase in that order instead of only cooling the axial modes, we observe a decrease in $p_{\uparrow\downarrow}+p_{\downarrow\uparrow}$ from $\approx$0.08 to $\approx$0.04.

\label{suppl_material}
\hfill \break
\bibliographystyle{apsrev4-2}
\bibliography{bibliography}
\clearpage

\end{document}